\documentclass[twocolumn,showpacs,amsmath,amssymb,prd,aps]{revtex4}
\usepackage{graphicx}
\usepackage{epstopdf}

\newcommand{\req}[1]{Eq.\,(\ref{#1})}

\begin{document} 
\title{%
Relic neutrinos: Physically consistent treatment of\\ effective number of neutrinos and neutrino mass 
%Cosmic Neutrino Spectrum:\\ Freezeout and chemical non-equilibrium%
%Massive Freestreaming Neutrinos: From Freezeout to Recombination%
%Massive relic neutrinos:  effective number of neutrinos from the end of BBN to end of recombination
}
\author{Jeremiah Birrell$^{1}$}
\author{Cheng-Tao Yang$^{2,3}$}
\author{Pisin Chen$^{2,3,4}$}
\author{Johann Rafelski$^{5}$}
\affiliation{%
$\quad$\\ 
{$^1$Program in Applied Mathematics, The University of Arizona, Tucson, Arizona, 85721, USA}}

\affiliation{%
$\quad$\\ 
{$^{2}$Department of Physics and Graduate Institute of Astrophysics, National Taiwan University, Taipei, Taiwan 10617}}
\affiliation{%
$\quad$\\ 
{$^{3}$Leung Center for Cosmology and Particle Astrophysics (LeCosPA),National Taiwan University, Taipei, Taiwan, 10617}}
\affiliation{%
$\quad$\\  
{$^{4}$Kavli Institute for Particle Astrophysics and Cosmology,SLAC National Accelerator Laboratory, Menlo Park, CA 94025, USA}}
\affiliation{%
$\quad$\\ 
{$^{5}$Department of Physics, The University of Arizona, Tucson, Arizona, 85721, USA}}
\date{December 10, 2013}% It is always \today, today,

\begin{abstract}
We perform a model independent study of the neutrino momentum distribution at freeze-out, treating the freeze-out temperature as a free parameter.  Our results imply that measurement of neutrino reheating, as characterized by the measurement of the effective number of neutrinos $N_\nu$, amounts to the determination of the neutrino kinetic freeze-out temperature within the context of the standard model of particle physics where the number of neutrino flavors is fixed and no other massless (fractional) particles arise.  At temperatures on the order of the neutrino mass, we show how cosmic background neutrino properties i.e. energy density, pressure, particle density, are modified in a physically consistent way as a function of neutrino mass and $N_\nu$. 
\end{abstract}

% 2010PACS, the Physics and Astronomy
%51.10.+y Kinetic and transport theory of gases (see also 05.20.Dd Kinetic theory in classical 
%95.30.Cq Elementary particle processes 
%14.60.Pq Neutrino mass and mixing 
%26.35.+c Big Bang nucleosynthesis (see also 98.80.Ft Origin, formation, and abundances of the elements in astronomy) 
%05.20.Dd Kinetic theory (see also 51.10.+y Kinetic and transport theory of gases) 
%13.15.+g Neutrino interactions 
%14.60.St Non-standard-model neutrinos, right-handed neutrinos, etc. 
%26.30.Jk Weak interaction and neutrino induced processes, galactic radioactivity 
%47.45.Ab Kinetic theory of gases 
%95.85.Ry Neutrino, muon, pion, and other elementary particles; cosmic rays 

\maketitle
%%%%%%%%%%%%%%%%%%%%%%%%%%%%%%%%%%%%%%%%%%%%%%
\section{Introduction}\label{Intro}
The  relic (i.e. background cosmic) neutrinos  have not been  directly measured~\cite{Ringwald:2009bg,Tupper:1987sf}. Their presence and properties are inferred from reaction dynamics throughout the history of the universe~\cite{Dolgov:2002wy,Lesgourgues:2006nd}. Important properties of the free-streaming relic neutrino background include
\begin{enumerate}
\item The number of neutrino flavors, $N_\nu^f=3$~\cite{aleph}.
\item Ratio of photon to neutrino temperature $T_\gamma/T_\nu$ (reheating ratio).\label{reheat}
\item Non-thermal distortions of the neutrino distribution which, in the present work, are captured by the neutrino fugacity $\Upsilon$. \label{non_thermal}
\item Neutrino handedness, including effects of mass captured in terms of  solution of  Einstein-Vlasov equation for massive neutrinos, see section~\ref{nu_mass_section}.
\end{enumerate}

For the purpose of computing the universe dynamics, all of these effects (excluding mass but including neutrino handedness) can be summarized as an effective number of neutrinos, a cosmological parameter defined for relativistic neutrinos by comparing the total neutrino energy density to the energy density of a massless fermion with two degrees of freedom and standard photon reheating ratio,
\begin{equation}\label{N_nu_c}
N^c_{\nu}\equiv\frac{\rho_\nu}{\frac{7}{120}\pi^2  \left(\left(\frac{4}{11}\right)^{1/3} T_\gamma\right)^4}.
\end{equation}
We emphasize that the cosmological effective number of neutrinos is distinct from the number of neutrino flavors, $N_\nu^f=3$, though the latter certainly would impact  the former should there be any doubt about the value of $N_\nu^f$. 

The standard reheating ratio $T_\gamma/T_\nu=(11/4)^{1/3}$ implied in \req{N_nu_c} arises from assuming that  the entropy from $e^\pm$ annihilation flows solely into photons. $N_\nu^c$ is normalized such that in the simplified model where there are no non-thermal distortions and standard reheating holds then $N_\nu^c=N_\nu^f=3$.  From now on, we will refer to $N_\nu^c$ simply as $N_\nu$ while $N_\nu^f$ will be taken to be the standard model value  of $N_\nu^f=3$ ~\cite{aleph}, and from now on will be  absorbed into the neutrino degeneracy factor.

As we will show, the non-integer number of neutrino degrees of freedom  $N_\nu$ reported experimentally~\cite{Planck} can be interpreted as an effect of neutrino freeze-out and reheating. This motivate a full reexamination of the neutrino freeze-out process employing the methods developed in the context of particle freeze-out in quark-gluon plasma hadronization~\cite{Letessier:2005qe,Kuznetsova:2006bh}. Our approach allows for us, in a model independent way, to relate the neutrino kinetic freeze-out temperature to $N_\nu$. 

Our model independent approach also allows us to derive  formulas for the neutrino energy density and pressure after freeze-out as functions of both the effective number of neutrinos and the neutrino masses.  These  allow for a self consistent study of the combined effects that a non-integer effective number of neutrinos and nonzero neutrino masses have on cosmological observables, a problem that, as discussed in \cite{Lesgourgues_book}, has proven difficult to approach.

The standard theory of neutrino freeze-out, based on the Einstein-Boltzmann equation with two body scattering and standard neutrino weak interactions \cite{Lopez:1998aq,Gnedin:1997vn,Mangano:2005cc}, calculates a (small) non-thermal distortion of the neutrino distribution after freeze-out. The current state of the art computation results in a slight deviation of $N_\nu$ from $3$ due to the participation of the high energy tail of the neutrino distribution in reheating, and hence a small entropy transfer from $e^\pm$ into neutrinos, together with the effect of neutrino oscillations, leading to $N_{\nu}^{\rm th}=3.046$ \cite{Mangano:2005cc}. 

Recent results from the nine-year WMAP observations, Table 7 of Ref.\cite{WMAP9}, and another independent study of BBN~\cite{Steigman:2012ve} favor an effective number of neutrinos at BBN of $N_\nu=3.55^{+0.49}_{-0.48}$ and $N_\nu=3.71^{+0.47}_{-0.45}$, respectively, while the newly released Planck data finds $N_\nu=3.30\pm 0.27$ \cite{Planck}.  However, this fit  produces a 2.5 s.d. tension with direct astrophysical measurements of the Hubble constant. Including priors from supernova surveys removes this tension and result in $N_\nu=3.62\pm 0.25$. 

There is currently a significant degree of  interest in the precise value of $N_\nu$, due to its impact on the spectrum of CMB fluctuations.  At the current level of precision,  it is certainly possible that the above measurements by these different methods agree, and agree with the theoretical result  $N_{\nu}^{\rm th}$. Though far from definitive, these results suggest the alternate possibility that some mechanism in addition to standard two body scattering leads to a greater entropy flow into neutrinos than predicted by standard weak interactions and hence a value of $N_\nu>3.046$.

Several scenarios for nonstandard neutrino interactions have been investigated, included neutrino electromagnetic properties \cite{Morgan1981247,Elmfors19973,Vogel1989,Fukugita1987,TEXONO2010,Broggini,Barranco2008,Hirsch2003,Tanimoto2000} and nonstandard neutrino electron coupling \cite{Mangano:2006}. In this paper we are not proposing a new mechanism for a modified neutrino freeze-out, but rather perform a model independent analysis of the impact of a delayed freeze-out on the neutrino momentum distribution -- motivated by the question of what precisely the measurement of $N_\nu=3.62\pm 0.25$ means for the neutrino momentum distribution.  

We work under the assumption that the increase in $N_\nu$ is due to the presence of conventional but not easily identified  interactions that keep neutrinos in equilibrium with the background $e^\pm$, $\gamma$ plasma down to a lower temperature.  In other words, we treat the kinetic freeze-out temperature, denoted by $T_k$, as a free parameter determined by the unknown physics and perform a parametric study of the dependence of the neutrino distribution on $T_k$. We show that a reduction in $T_k$, by whatever mechanism, leads to an increase in $N_\nu$ and is capable of achieving the values seen in the Planck data.

There are two physical effects which combine in our analysis to yield to the end result:
\begin{enumerate}
\item 
Chemical freeze-out, $T_{ch}$, the temperature at which particle number changing processes such as $e^+e^-\longleftrightarrow \nu\bar\nu$ effectively cease, and kinetic freeze-out, the temperature $T_k$ at which all momentum exchanging processes such as $e^\pm \nu\longleftrightarrow e^\pm \nu$ cease, are distinct.  Once the universe temperature drops below  the chemical freeze-out temperature  $T_{ch}$, there are no reactions that, in a noteworthy fashion, can change the neutrino abundance and so particle number is conserved. However, the distribution remains in kinetic equilibrium, and hence exchanges momentum with $e^\pm$, down until $T=T_k$.
\item 
Our effect requires that the temperature interval $T_k<T<T_{ch}$  overlaps with  $T\approx m_e$ when the electron-positron mass becomes a significant scale and reheating occurs.  This allows annihilation of $e^\pm$ to feed energy and entropy into neutrinos and reduce the  photon-neutrino temperature ratio $T_\gamma/T_\nu$. As we will see, the freeze-out temperature for standard model neutrino scattering processes is on the border of this regime.
\end{enumerate}

That `neutrino reheating' leads to an increase in $N_\nu$ is quite well known~\cite{Lopez:1998aq,Gnedin:1997vn,Mangano:2005cc}, as it is precisely this effect that leads to the standard value of $N_\nu=3.046$. However, it is not well known  that   reheating of neutrinos is  accompanied by an underpopulation of neutrino phase space relative to an equilibrium distribution. This underpopulation is characterized in the present context by a little-known cosmological model parameter, the neutrino fugacity  $\Upsilon_\nu$. Its significance for neutrino cosmology has been previously recognized \cite{Bernstein:1985,Dolgov:1993}  but is not widely appreciated. On the other hand, in other physical processes that involve decoupling and freeze-out such chemical parameters are in daily use as already noted~\cite{Letessier:2005qe}.

Since we ask how the kinetic freeze-out $T_k$ needs to be modified in order to explain a given value $N_\nu$, in principle one can wonder if $T_{ch}$ should also change. The general experience from other areas of physics is that it is much more difficult to find changes in $T_{ch}$ beyond two body interaction processes. The reason that  $T_k$ is more easily modified is the possible appearance of collective coherent scattering processes of the type neutrino-pasmon scattering which add to elastic scattering and thus alter $T_k$ but normally vanish in particle changing processes, leaving $T_{ch}$ unchanged. Therefore, in our analysis, we consider the chemical freeze-out process to be fixed by standard model weak interactions. In addition to the above motivation, even if  $T_{ch}$ were modified, its precise value is entirely immaterial to the present study as long as $T_{ch}$ occurs before $e^\pm$ annihilation begins in earnest, as demonstrated in \cite{Birrell:2013}. Under this assumption, we examine the effect that a nonstandard neutrino kinetic freeze-out temperature $T_k$ has on the form of the cosmic neutrino distribution and effective number of neutrinos after decoupling  in a model independent fashion (i.e. treating $T_k$ as a free parameter). 

In section \ref{neqnu}  we discuss the general form of the non-equilibrium neutrino distribution, including the significance of the fugacity parameter. In \ref{EVEq} we derive the form of the free-streaming neutrino distribution using the Einstein-Vlasov equation.  In \ref{stress_energy}   we compute various moments of the distribution.  In section  \ref{Tnugam} we compute the relation between neutrino fugacity, the reheating temperature ratio, and the kinetic freeze-out temperature.  In section \ref{N_eff_section} we discuss the impact on the effective number of neutrinos.  We discuss the combined impact of neutrino mass and fugacity parameters when the temperature is on the order of the neutrino mass in section \ref{nu_mass_section}. We present our conclusions and discussion in section~\ref{discuss}.

%%%%%%%%%%%%%%%%%%%%%%%%%%%%%%%%%%%%%%%%%%%%%%%%%%%%%%%%%%%%%
\section{Nonequilibrium neutrinos}\label{neqnu}
\subsection{Chemical and Kinetic Equilibrium }\label{EVEq}
Prior to the neutrino chemical freeze-out temperature, $T_{ch}$, number changing processes are significant and keep neutrinos in chemical (and thermal) equilibrium, implying that the distribution function of each neutrino flavor has the Fermi-Dirac form, obtained by maximizing entropy at fixed energy
\begin{equation}\label{equilibrium}
f_{c}(t,E)=\frac{1}{\exp(E/T)+1}, \text{ for } T> T_{ch}.
\end{equation}
 When $T_k<T<T_{ch}$, number changing process  no longer occur rapidly enough to keep the distribution in chemical equilibrium but there is still sufficient momentum exchange to keep the distribution in thermal equilibrium.  The distribution function is therefore obtained by maximizing entropy, with a fixed energy, particle number, and antiparticle number separately,  implying that the distribution function function has the form
\begin{equation}\label{kinetic_equilib}
f_k(T,E)=\frac{1}{\Upsilon_\nu^{-1}\exp(E/T)+1}, \text{ for }T_k< T< T_{ch}.
\end{equation}

The fugacity, $\Upsilon_\nu\equiv e^\sigma$, controls the occupancy of phase space and is necessary once $T<T_{ch}$ in order to conserve particle number.  The effect of $\sigma$ is similar after that of chemical potential $\mu$, except that $\sigma$ is equal for particles and antiparticles, and not opposite, as noted in  \cite{Bernstein:1985,Dolgov:1993}.  This means $\sigma>0$ ($\Upsilon_\nu>1$) increases the density of both particles and antiparticles, rather than increasing one and decreasing the other as is common when the chemical potential is associated with conservation laws such as lepton number.  Similarly, $\sigma<0$ $(\Upsilon_\nu<1)$ decreases both. The fact that $\sigma$ is not opposite for particles and antiparticles reflects the fact that both  the number of particles and the number of antiparticles are conserved after chemical freeze-out, and not just their difference.  The equality reflects the fact that any process that modifies  the distribution would affect both particle and antiparticle distributions in the same fashion. 

The use of $\Upsilon_\nu$ to account for the processes that feed into neutrinos is nearly exact in the temperature interval after chemical and before kinetic freeze-out, since scattering processes re-equilibrate the momentum distribution to this shape in order to maximize the entropy content. However, it is an approximation when the additional particle feeding occurs  near kinetic freeze-out, where the energy dependence of the neutrino cross sections becomes significant, leading to an energy dependent freeze-out and therefore additional non-thermal distortions.  These could be thought of as allowing $\Upsilon_\nu$ to be momentum dependent. In the particular case investigated in ~\cite{Mangano:2005cc}, these non-thermal distortions are small, below 5\%.  In this work, we will restrict our attention to the simplified model of a momentum independent $\Upsilon_\nu$.

%%%%%%%%%%%%%%%%%%%%%%%%%%%%%%%%%%%%%%%%%%%%%%%
\subsection{Einstein-Vlasov Equation in FRW Spacetime}\label{EVEq}
 We begin our analysis with the simplest regime (from the neutrino perspective), $T<T_k$, when both number changing and momentum exchanging interactions have ceased and neutrinos begin to freely stream. The general relativistic Boltzmann equation describes the dynamics of a gas of particles that travel freely in between point interactions in an arbitrary spacetime~\cite{Andreasson:2011ng,cercignani,bruhat,ehlers} 
\begin{equation}\label{boltzmann}
p^\alpha\partial_{x^\alpha}f-\Gamma^j_{\mu\nu}p^\mu p^\nu\partial_{p^j}f=C[f].
\end{equation}
Here $ \Gamma^\alpha_{\mu\nu}$ is the affine connection (Christoffel symbol),  $f$ is a function on the mass shell \begin{equation}
g_{\alpha\beta}p^\alpha p^\beta=m^2,
\end{equation}
hence Greek indices are summed from $0$ to $3$ whereas $j$ is only summed from $1$ to $3$.  When collisions are negligible, such as for $T<T_k$, we have $C[f]=0$ and all particles  move on geodesics, yielding the Einstein-Vlasov equation.

We now specialize to collision free homogeneous isotropic cosmological solutions and therefore assume the flat FRW ansatz for the spacetime metric
\begin{equation}
g=dt^2-a(t)^2(dx^2+dy^2+dz^2).
\end{equation}
We will make the simplifying assumption of a perfectly homogeneous universe. See \cite{Wong} for a review of the results and challenges associated with the study of inhomogeneities.

  Due to homogeneity and isotropy, the neutrino distribution function depends on $t$ and $p^0=E$ only.  Therefore \req{boltzmann} becomes
\begin{equation}\label{VEeqFLR}
E\partial_tf+(m^2-E^2)\frac{\partial_ta}{a}\partial_{E}f=0.
\end{equation}
The general solution to \req{VEeqFLR} is known.  For example, see ~\cite{bruhat} or \cite{Wong}.
\begin{equation}\label{general_sol}
f(t,E)=K(x),\hspace{3mm} x=\frac{a(t)^2}{D^2}(E^2-m^2),
\end{equation}
where $K$ is an arbitrary smooth function and $D$ is an arbitrary constant with units of mass.  To continue the evolution beyond the freeze out time, $t_k$, we must choose $K$ to match at $t_k$ the equilibrium distribution \req{equilibrium}.

With this in mind, we let 
\begin{equation}\label{K_func}
K(x)=\frac{1}{\Upsilon_\nu^{-1}e^{\sqrt{x+m^2/T_k^2}}+ 1}
\end{equation}
and $D=T_k a(t_k)$ to match \req{kinetic_equilib} at freeze-out. The Fermi-Dirac-Einstein-Vlasov (FDEV) distribution function for neutrinos after freeze-out is then
\begin{equation}\label{neutrino_dist}
f(t,E)=\frac{1}{\Upsilon_\nu^{-1}e^{\sqrt{(E^2-m^2)/T_\nu^2+m_\nu^2 /T_k^2}}+ 1}
\end{equation}
where 
\begin{equation}\label{Tneutrino_dist}
T_\nu(t)=\frac{T_ka(t_k)}{a(t)}.  
\end{equation}
\req{neutrino_dist} provides the distribution function that describes a gas of neutrinos that have been free streaming in an expanding universe since they froze out at $T_\nu(t_k)=T_k$. We will call $T_\nu$ in \req{Tneutrino_dist} the neutrino background temperature, even though  the distribution of free streaming particles has a thermal shape only for $m=0$.   This language is, however, reasonable since  apart from the reheating factor of photons due to $e^+e^-$ annihilation, which we discuss in section \ref{Tnugam}, $T_\nu$ tracks the photon background temperature. 

%%%%%%%%%%%%%%%%%%%%%%%%%%%%%%%%%%%%%%%%%%%%%%%
\subsection{Moments of FDEV Distribution}\label{stress_energy}
Here we compute the stress energy tensor, number current, and entropy current associated with the distribution \req{neutrino_dist}
\begin{align}
\label{Tmndef}{\cal T}^{\mu\nu}&=\frac{g_\nu}{8\pi^3}\int f\frac{p^\mu p^\nu}{p_0}\sqrt{-g}d^3p,\\
\label{nmdef} n^\nu&=\frac{g_\nu}{8\pi^3}\int f \frac{p^\nu}{p_0} \sqrt{-g}d^3p,\\
\label{smdef} s^\mu&=-\frac{g_\nu}{8\pi^3}\int h\frac{p^\mu}{p_0}\sqrt{-g}d^3p,\\
h&=f\ln(f)+(1-f)\ln(1-f)\notag
\end{align}
where $g_\nu$ is the neutrino degeneracy (not to be confused with the metric factor $\sqrt{-g}=a^3$).  We first work with the general form of $f$ given in \req{general_sol} and later specialize to the explicit form \req{neutrino_dist}. 

Isotropy of the metric and of $f$ in momentum space implies that the off diagonal elements of the stress energy tensor and spacial components of the particle number and entropy currents vanish and that the pressure is isotropic.  Hence we must compute 
\begin{align}
{\cal T}^{00}&=a^3\frac{g_\nu}{8\pi^3}\int fE d^3p,\\
{\cal T}^{ii}&=\frac{1}{3}a^3\frac{g_\nu}{8\pi^3}\int f \frac{|p|^2}{E} d^3p, \hspace{2mm} i=1...3,\\
n^{0}&=a^3\frac{g_\nu}{8\pi^3}\int f d^3p,\\
s^0&=-a^3\frac{g_\nu}{8\pi^3}\int h d^3p
\end{align}
where $|p|$ is the Euclidean norm of the spacial components of $p^\mu$ and $E=p^0$ is given by
\begin{equation}
m_\nu^2=E^2-a(t)^2|p|^2.
\end{equation}
Computing ${\cal T}^{00}$ we find
\begin{align}
{\cal T}^{00}&=\frac{g_\nu a^3}{2\pi^2}\int_0^\infty K((E^2-m_\nu^2)/T_\nu^2)E|p|^2d|p|\notag\\
&= \frac{g_\nu a^3}{2\pi^2}\int_0^\infty K(a^2 p^2/T_\nu^2)(m_\nu^2+a^2p^2)^{1/2}|p|^2d|p|\notag\\
&=\frac{g_\nu}{2\pi^2}\int_0^\infty K(z^2/T_\nu^2)(m_\nu^2+z^2)^{1/2}z^2dz
\end{align}
where we made a change of variables $z=a(t)|p|$.  Note that $z$ is the physically measured momentum.  Similarly
\begin{align}
{\cal T}^{ii}&=\frac{g_\nu}{6\pi^2a^2}\!\int_0^\infty\!\!\! K(z^2/T_\nu^2)(m_\nu^2+z^2)^{-1/2}z^4dz,\\
 n^0&=\frac{g_\nu}{2\pi^2}\!\int_0^\infty\!\!\! K(z^2/T_\nu^2)z^2 dz,\\
s^0&=-\frac{g_\nu}{2\pi^2}\!\int_0^\infty\!\!\!H(z^2/T_\nu^2)z^2dz,\\ 
H&=K\ln K +(1-K)\ln(1-K).\label{entropy_integrand}
\end{align}

We now rename $z$ to $p$, so that $p$ represents the magnitude of the physical momentum, drop the superscripts, and insert \req{K_func} for $K$, giving the energy density, pressure, and number density for each neutrino flavor
\begin{align}
\rho&=\frac{g_\nu}{2\pi^2}\!\int_0^\infty\!\!\!\frac{\left(m_\nu^2+p^2\right)^{1/2}p^2dp }{\Upsilon_\nu^{-1}e^{\sqrt{p^2/T_\nu^2+m_\nu^2/T_k^2}}+ 1},\label{neutrino_rho}\\[0.2cm]
P&=\frac{g_\nu}{6\pi^2}\!\int_0^\infty\!\!\!\frac{\left(m_\nu^2+p^2\right)^{-1/2}p^4dp }{\Upsilon_\nu^{-1} e^{\sqrt{p^2/T_\nu^2+m_\nu^2/T_k^2}}+ 1},\label{neutrino_P}\\[0.2cm]
n&=\frac{g_\nu}{2\pi^2}\!\int_0^\infty\!\!\!\frac{p^2dp }{\Upsilon_\nu^{-1}e^{\sqrt{p^2/T_\nu^2+m_\nu^2/T_k^2}}+ 1}.
\label{num_density}
\end{align}
These differ from the corresponding expressions for an equilibrium distribution in Minkowski space by the replacement $m\rightarrow m T_\nu(t)/T_k$  {\em only} in the exponential.  

By making a change of variables $u=p/T_\nu$, one sees that both $n$ and $s$ are proportional to $T_\nu^3$.  By definition, $T_\nu$ is inversely proportional to $a$, hence
\begin{equation}\label{const_entropy}
a^3n=\text{constant}\text{ and } a^3s=\text{constant}.
\end{equation}
This proves that the particle number and entropy in a comoving volume are conserved.  We emphasize that this result does not depend on the particular form of $K$ that defines the shape of the momentum distribution at freeze-out. Note further that since an eV scale or below neutrino mass is at least 6 orders of magnitude smaller than $T_k$, we can ignore the neutrino mass in the exponent. This remark does not extend to the energy factor multiplying the FD-distribution in the calculation of energy density or pressure and therefore \req{neutrino_rho} and \req{neutrino_P} have unusual properties, while \req{num_density} is just what one may naively expect, remembering that the mass term in this expression is completely negligible.

%%%%%%%%%%%%%%%%%%%%%%%%%%%%%%%%%%%%%%%%%%%%%%%%%%%%%%%%%%%%%%%%%%%%%%%%%%%%%%%%%%%%%%%%%%
\section{Neutrino Fugacity and Photon to Neutrino Temperature Ratio}\label{Tnugam}
To complete our characterization of the neutrino distribution as it would have evolved from decoupling until recombination, we must understand the relation between the  the kinetic freeze-out temperature, photon to neutrino temperature ratio, and the fugacity. To that end, we now focus on the regime between chemical and kinetic freeze-out, $T_k<T<T_{ch}$. 

Before the reheating period, neutrinos, photons, electrons, and positrons have the same temperature.  When the temperature approaches and drops below the electron mass, the electrons and positrons annihilate.   After photon reheating, both the neutrino and photon temperatures evolve inversely proportional to $a$, so their ratio after reheating equals their ratio today.   The resulting ratio of photon to  neutrino temperatures  has often been studied, but not the type of model independent study of the $T_k$ parameter space that we present here. In the following we  use subscripts $1$ and $2$ to denote quantities before and after reheating, respectively.

We first outline the physics of the situation qualitatively. For $T_k<T<T_{ch}$, the evolution of the temperature of the common $e^\pm,\gamma,\nu$ plasma and the neutrino fugacity are determined by conservation of comoving neutrino number (since $T<T_{ch}$) and conservation of entropy.  The latter condition is not exactly correct once one drops the assumption of an instantaneous chemical freeze-out, but it is a very good approximation as shown in \cite{Birrell:2013}.  As shown in section \ref{neqnu}, after thermal freeze-out the neutrinos begin to free-stream and therefore $\Upsilon_\nu$ is constant, the neutrino temperature evolves as $1/a$, and the comoving neutrino entropy and neutrino number are exactly conserved \req{const_entropy}.  The photon temperature then evolves to conserve the comoving entropy in photons, electrons, and positrons.  As annihilation occurs, entropy from $e^+e^-$ is fed into photons, leading to reheating.  We now make this analysis quantitative in order to derive a relation between the reheating temperature ratio and neutrino fugacity.

When the (common) temperature $T_1$ is much larger than the electron mass and $T_k$, the entropy in a given comoving volume, $V_1$, is the sum of relativistic neutrinos (with $\Upsilon_\nu=1$), electrons, positrons, and photons
\begin{equation}
S(T_1)=\left(\frac{7}{8}g_\nu+\frac{7}{8}g_{e^\pm} +g_\gamma \right)\frac{2\pi^2}{45} T_1^3V_1
\end{equation}
where $T_1$ is the common neutrino, $e^+e^-$, and $\gamma$ temperature. The number of neutrinos and anti-neutrinos in this same volume is
\begin{equation}
\mathcal{N}_\nu(T_1)=\frac{3g_\nu}{4\pi^2}\zeta(3)T_1^3V_1.
\end{equation}
The particle-antiparticle, flavor, and spin-helicity statistical factors are $g_\nu=6$, $g_{e^\pm}=4$, $g_\gamma=2$.

As discussed above, distinct chemical and thermal freeze-out temperatures lead to a non-equilibrium modification of the neutrino distribution in the form of a fugacity factor $\Upsilon_\nu$.  This leads to the following expressions for neutrino entropy and number at $T=T_k$ in the comoving volume
\begin{align}
S(T_k)=&\left(\frac{2\pi^2}{45}g_\gamma T_k^3+S_{e^\pm}(T_k)+S_{\nu}(T_k)\right)V_k,\\
\mathcal{N}_{\nu}(T_k)=&\frac{g_\nu}{2\pi^2}\int_0^\infty \frac{u^2 du}{\Upsilon_\nu^{-1}(T_k)e^u+1}T_k^3V_k.
\end{align}

After neutrino kinetic freeze-out and when $T\ll m_e$, i.e. after reheating has completed and almost all of the $e^+e^-$ have annihilated, the entropy in neutrinos is conserved independently of the other particle species, the electron entropy is negligible, and the photon entropy is 
\begin{equation}
S_{\gamma}(T_2)=\frac{2 \pi^2}{45}g_\gamma T_{\gamma,2}^3 V_2.
\end{equation}
 Note that we must now distinguish between the neutrino and photon temperatures.\\

Conservation arguments give the following three relations.
\begin{enumerate}
\item Conservation of comoving neutrino number:
\begin{equation}\label{eq_1}
\frac{T_1^3V_1}{T_k^3V_k}=\frac{2}{3\zeta(3)}\int_0^\infty \frac{u^2 du}{\Upsilon_\nu^{-1}(T_k)e^u+1}.
\end{equation}
\item Conservation of $e^\pm$, $\gamma$, neutrino entropy before neutrino freeze-out:
\begin{align}\label{eq_2}
&\left(\frac{7}{8}g_\nu+\frac{7}{8}g_{e^\pm} +g_\gamma \right)\frac{2\pi^2}{45} T_1^3V_1=\\
&\left(S_{\nu}(T_k)+S_{e^\pm}(T_k)+\frac{2\pi^2}{45}g_\gamma T_k^3\right)V_k.\notag
\end{align}
\item Conservation of $e^\pm$, $\gamma$ entropy after neutrino freeze-out  (at this point, neutrino entropy is conserved independently):
\begin{equation}\label{eq_3}
\frac{2 \pi^2}{45}g_\gamma T_{\gamma,2}^3 V_2=\left(\frac{2\pi^2}{45}g_\gamma T_k^3+S_{e^\pm}(T_k)\right)V_k.
\end{equation}
\end{enumerate}
These relations allow us to solve for the fugacity, reheating ratio, and effective number of neutrinos in terms of the freeze-out temperature, irrespective of the details of the dynamics that leads to a particular freeze-out temperature.  

%%%%%%%%%%%%%%%%%%%%%%%%%%%%%%%%%%%%%%%%%%%%%%%%%%%%%
\subsection{Neutrino Fugacity}\label{neutrino_fugacity}
\begin{figure} 
\centerline{\includegraphics[height=6.4cm]{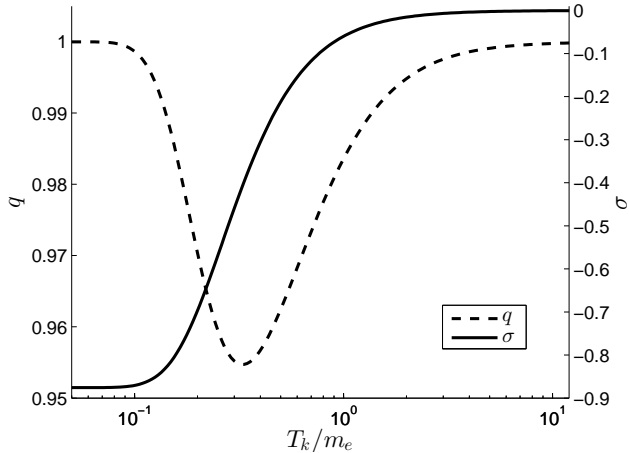}}
\caption{Deceleration parameter (left axis) and log of neutrino fugacity (right axis) as  functions of kinetic freeze-out temperature.}\label{fig:Upsilon_q}
 \end{figure}

For $T\gg m_e$ the universe is radiation dominated and $T\propto 1/a$.  Therefore we can normalize the scale factor so that $T_1^3V_1\rightarrow 1$ as $T_1\rightarrow\infty$. With this normalization, \req{eq_1} and \req{eq_2} become two equations that can be solved numerically for $\Upsilon_\nu(T_k)$, shown in figure \ref{fig:Upsilon_q},  and $a(T_k)=V_k^{1/3}$. We emphasize that $\Upsilon_\nu\neq 1$ is an unavoidable consequence of the freeze-out process, whenever the interval $T_k<T<T_{ch}$ contains temperatures on the order of the electron mass.  This latter condition is critical.  If freeze-out occurs while $e^\pm$ are still effectively massless then, after setting $m/T_k=0$ in \req{eq_2}, we see that $T_1^3V_1=T_k^3V_k$ i.e. the temperature evolves as $T=1/a$.  Inserting this into \req{eq_1} then implies that $\Upsilon_\nu=1$.  This behavior is seen in figure  \ref{fig:Upsilon_q} when $T_k/m_e$ is large.  When $T_k/m_e=O(1)$ this argument no longer holds; there is no longer any solution with $\Upsilon_\nu=1$. From figure \ref{fig:Upsilon_q} we see $\Upsilon_\nu$ is monotonically decreasing with $T_k$, indicating an underpopulation of phase space compared to equilibrium.

Figure \ref{fig:Upsilon_q}  also shows the deceleration parameter
\begin{equation}
q\equiv-\frac{\ddot{a}a}{\dot{a}^2}=\frac{1}{2}\left(1+3\frac{P}{\rho}\right).
\end{equation}
For a purely radiation dominated universe $q=1$.   A mass scale that becomes relevant at a particular temperature causes $q$ to drop below unity, towards the matter dominated value $q=1/2$.  $q<1$ is therefore an indicator that conditions are right for $\Upsilon$ to be pushed off of its equilibrium value of unity if in addition we have $T_k<T<T_{ch}$, per our discussion above.  A similar plot was obtained in Ref.\cite{Birrell:2013} for a more detailed model of a smooth (i.e. not instantaneous) chemical freeze-out, where we showed that the instantaneous freeze-out approximation has error of less than  $1\%$.  The result of the argument given here is equivalent to the entropy conserving curve in that figure in Ref.\cite{Birrell:2013}. Considering the current neutrino mass bounds of $O(0.1)$ eV,  the values of $\sigma$ achievable in a delayed freeze-out scenario are comparable to $m_\nu/T_{r}$ where $T_r=0.253$ eV is the recombination temperature.  This suggests that the effects of $\sigma$  may compete with the impact of neutrino mass. We will discuss this further in section \ref{nu_mass_section}.

%%%%%%%%%%%%%%%%%%%%%%%%%%%%%%%%%%%%%%%
%%%%%%%%%%%%%%%%%%%%%%%%%%%%%%%%%%%%%%%

%%%%%%%%%%%%%%%%%%%%%%%%%%%%%%%%%%%%%%%%%%%%%%%%%%%%%
\subsection{Reheating Ratio}\label{reheat_ratio}
We now derive the relation between the reheating temperature ratio and neutrino fugacity. Using \req{eq_2} and \req{eq_3} we can eliminate $S_{e^\pm}(T_k)$ and obtain
\begin{align}\label{reheat_ratio_eq_1}
 &\left(\frac{7}{8}g_\nu+\frac{7}{8}g_{e^\pm} +g_\gamma \right)\frac{2\pi^2}{45} T_1^3V_1\\
 &-S_\nu(T_k)V_k= \frac{2 \pi^2}{45}g_\gamma T_{\gamma,2}^3 V_2.\notag
\end{align}
Note that this by no means implies that the entropy remaining in $e^+e^-$ at freeze-out plays no role in our discussion; it was crucial for computing $\Upsilon_\nu(T_k)$, shown in figure \ref{fig:Upsilon_q}. 

Dividing both sides of \req{reheat_ratio_eq_1} by $\frac{2 \pi^2}{45}g_\gamma T_k^3V_k$ and using \req{eq_1} we find
\begin{align}
&\frac{2}{3\zeta(3)}\left(1+\frac{7}{8}\frac{g_\nu+g_{e^\pm}}{g_\gamma}\right)\int_0^\infty \frac{u^2 du}{\Upsilon_\nu^{-1}(T_k)e^u+1}\label{eq_4}\\
&-\frac{45}{2\pi^2g_\gamma}S_\nu(T_k)/T_k^3=\frac{T_{\gamma,2}^3 V_2}{T_k^3V_k}.\notag
\end{align}

 From \req{Tneutrino_dist}, the neutrino temperature after kinetic freeze-out is 
\begin{equation}
T_{\nu,2}=\frac{a(t_k)T_k}{a(t_2)}=\left(\frac{V_kT_k^3}{V_2}\right)^{1/3}.
\end{equation}
Therefore \req{eq_4} gives the photon to neutrino temperature ratio after freeze-out as a function of the neutrino fugacity
\begin{align}
&\frac{2}{3\zeta(3)}\left(1+\frac{7}{8}\frac{g_\nu+g_{e^\pm}}{g_\gamma}\right)\int_0^\infty \frac{u^2 du}{\Upsilon_\nu^{-1}e^u+1}\label{eq_5}\\
&-\frac{45}{2\pi^2g_\gamma}S_\nu(T_k)/T_k^3=\left(\frac{T_\gamma}{T_\nu}\right)^3.\notag
\end{align}

We emphasize that $S_\nu(T_k)$ scales as $T_k^3$ and so $S_\nu(T_k)/T_k^3$ depends only on $\Upsilon_\nu$.  We now write $\Upsilon_\nu=e^\sigma$ and Taylor expand $\log\left(T_\gamma/T_\nu\right)$ about $\sigma=0$ to obtain
\begin{align}\label{Upsilon_ratio}
\frac{T_\gamma}{T_\nu}&=a\Upsilon_\nu^{b}\left(1+c\sigma^2+O(\sigma^3)\right),\\
\label{value_a}
a&=\left(1+\frac{7}{8}\frac{g_{e^\pm}}{g_\gamma}\right)^{1/3}=\left(\frac{11}{4}\right)^{1/3}\approx 1.4010,\\
\label{value_b}
b&=\frac{\pi^2}{27\zeta(3)}\frac{1+\frac{7}{8}\frac{g_\nu+g_{e^\pm}}{g_\gamma}-\frac{3645}{8\pi^6}\zeta(3)^2\frac{g_\nu}{g_\gamma}}{1+\frac{7}{8}\frac{g_{e^\pm}}{g_\gamma}}\\
&\approx 0.367,\\
c&\approx -0.0209.
\end{align}
The second order coefficient, c, is significantly smaller than than $b$, making the power law approximation very accurate (to within $2\%$ relative error) in the region of interest $.4\leq\Upsilon_\nu\leq 1$.  For additional precision we have included the second order term as well, bringing the relative error down to less than $5\times 10^{-4}$ over the same range of $\Upsilon$.  

The above analytic discussion presents another perspective on the power law obtained in \cite{Birrell:2013} using the more complex model of a smooth chemical freeze-out.  There, we found the nearly identical relation 
\begin{equation}
\Upsilon_\nu=0.420\left(\frac{T_\gamma}{T_\nu}\right)^{2.57}
\end{equation}
to within $1\%$ over the region $.4\leq\Upsilon_\nu\leq 1$ by a numerical fitting procedure rather than an analytic argument. For comparison with parameters $a, b$ in \req{value_a} and \req{value_b}, this numerical approximation translates to
\begin{equation}
\frac{T_\gamma}{T_\nu}=\tilde{a}\Upsilon_\nu^{\tilde{b}}, \hspace{2mm} \tilde{a}\approx 1.4015, \hspace{2mm} \tilde{b}\approx 0.389.
\end{equation}

%%%%%%%%%%%%%%%%%%%%%%%%%%%%%%%%%%%%%%%
\subsection{Effective number of neutrinos}\label{N_eff_section}
For (effectively) massless neutrinos, a deviation of the distribution function from the equilibrium form with standard reheating is summarized by the effective number of neutrinos, $N_\nu$, defined in \req{N_nu_c}.  As discussed above, the currently accepted theoretical value is $N_{\nu}=3.046$ \cite{Mangano:2005cc} after reheating, while Planck data gives $N_{\nu}=3.36\pm0.34$ (CMB only) and $N_{\nu}=3.62\pm0.25$ ($\text{CMB} + H_0$) \cite{Planck}.\\

%%%%%%%%%%%%%%%%%%%%%%%%%%%%%%%%%%%%%%%
\begin{figure}[h]
\centerline{\includegraphics[height=6.4cm]{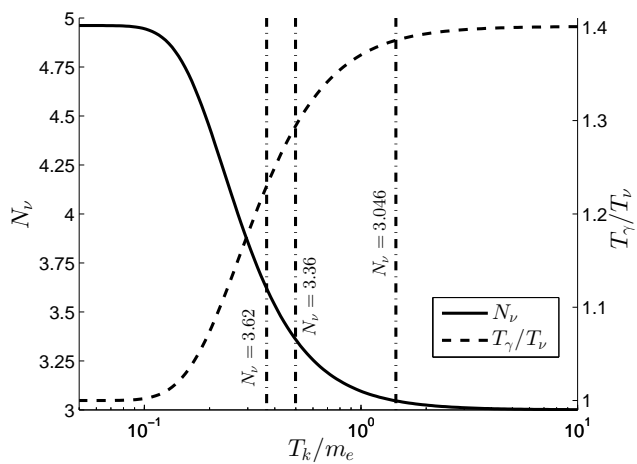}}
\caption{Effective number of neutrinos and photon to neutrino temperature ratio after reheating, both as a functions of $T_k$.}\label{fig:N_eff}
 \end{figure}
%%%%%%%%%%%%%%%%%%%%%%%%%%%%%%%%%%%%%%%
%%%%%%%%%%%%%%%%%%%%%%%%%%%%%%%%%%%%%%%
\begin{figure}[h]
\centerline{\includegraphics[height=6.4cm]{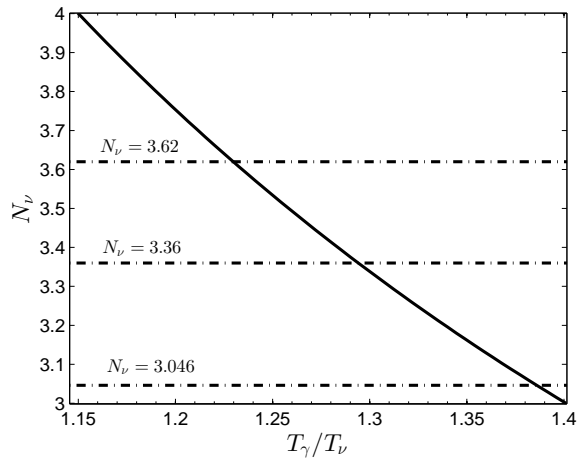}}
\caption{ $N_\nu$ after reheating, as a function of photon to neutrino temperature ratio.}\label{fig:reheat_ratio}
 \end{figure}
%%%%%%%%%%%%%%%%%%%%%%%%%%%%%%%%%%%%%%%

After reheating, both $T_\nu$ and $T_\gamma$ evolve inversely proportional to the scale factor, and so the reheating ratio remains constant. Combining this with the fact that for massless neutrinos, $\rho_\nu$ is proportional to $T_\nu^4$ implies $N_\nu=$ constant after reheating, at least until the temperature reaches the neutrino mass scale, at which point the definition \req{N_nu_c} becomes inappropriate for characterizing the number of massless degrees of freedom.  Even after the neutrino mass scale does become relevant, we will still use $N_\nu$ to refer to the value of the effective number of neutrinos that was established at freeze-out, even though the relation \req{N_nu_c} will no longer hold.

  Using \req{Upsilon_ratio} and $\Upsilon_\nu(T_k)$ from figure \ref{fig:Upsilon_q} we obtain $T_\gamma/T_\nu$ and $N_\nu$ after reheating as a function of $T_k$.  Most importantly, note that the decrease in the reheating ratio is able to overcome the drop in phase space occupancy $\Upsilon_\nu<1$, the combined effect being an increase in $N_\nu$ as shown in figure \ref{fig:N_eff}.  It saturates for small $T_k$ since both $\Upsilon_\nu$ and the reheating ratio do.

We see that the effect of a delayed freeze-out is capable of matching a non-integer number of neutrinos in the range that is currently favored by Planck. In figure \ref{fig:N_eff}  the vertical lines  indicate the value of the freeze-out temperature that corresponds to the indicated value of $N_\nu$.  A measurement of $N_\nu$ therefore is here demonstrated as being a measurement of the kinetic freeze-out temperature.  Moreover, the measurement of $N_\nu$  also determines the reheating temperature ratio between photons and neutrinos, shown in the solid line in figure \ref{fig:reheat_ratio}. Here the horizontal lines guide the eye.

%%%%%%%%%%%%%%%%%%%%%%%%%%%%%%%%%%%%%%%
\begin{figure}[h]
\centerline{\includegraphics[height=6.4cm]{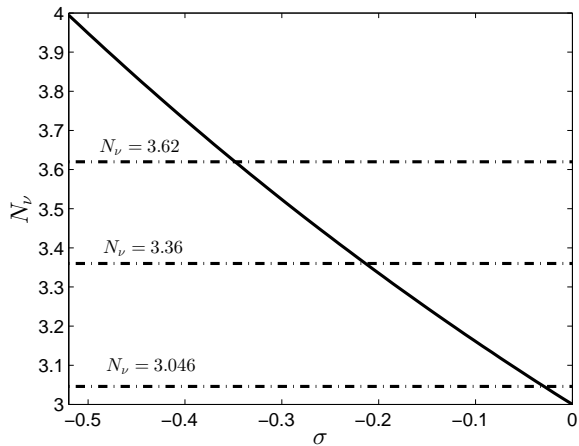}}
\caption{Effective number of neutrinos after reheating, as a function of $\sigma$.}\label{fig:N_sigma}
 \end{figure}
%%%%%%%%%%%%%%%%%%%%%%%%%%%%%%%%%%%%%%%
Figure \ref{fig:N_sigma} shows the effective number of neutrinos after reheating, as a function of $\sigma=\ln \Upsilon$.  Using the second order expansion for the reheating ratio in \req{Upsilon_ratio}, we can also present the analytic formula
\begin{equation}\label{N_nu_approx}
N_\nu=\frac{360}{7\pi^4}\frac{e^{-4b\sigma}}{(1+c\sigma^2)^4}\int_0^\infty \frac{u^3}{e^{u-\sigma}+1}du\left(1+O(\sigma^3)\right).
\end{equation}
where $a,b,c$ are the same as in \req{Upsilon_ratio}. The relative error of this approximation is less than $0.002$ over the range $-0.9\leq\sigma\leq 0$.  The second order Taylor expansion of the integral in this expression is not sufficiently accurate over the desired range of $\sigma$, so we leave it in the presented integral-analytic form.  For applications, it must be evaluated numerically. 

%%%%%%%%%%%%%%%%%%%%%%%%%%%%%%%%%%%%%%%%%%%%%%%%%%%%%%%%%%%%%%%%%%
\section{Neutrino Mass and Fugacity in Ion Recombination Era}\label{nu_mass_section}
To this point, we have characterized the freeze-out process in terms of the kinetic freeze-out temperature $T_k$ and  obtained the form of neutrino momentum distribution that results.   In section~\ref{EVEq} we  presented a solution of  the free streaming dynamics of neutrinos which allows us to obtain the form of neutrino distribution at any later epoch.

The precise form of the neutrino distribution and in particular the fugacity parameter $\Upsilon_\nu=e^\sigma$ we introduced in this paper become physically relevant when the Universe temperature approaches and drops below the neutrino mass. This is so since as long as the neutrino mass is negligible, the shape of the neutrino distribution as determined in this work has no impact on expansion dynamics of the Universe considering that, in the absence of a scale parameter, there is no modification of the equation of state. We now characterize the regime when the mass of neutrinos become a relevant scale  including the effects of fugacity $\Upsilon_\nu$.

To  compare the energy density \req{neutrino_rho} and pressure \req{neutrino_P} to that of a massless particle distribution with $\Upsilon=1$, we make a change of variables $u=p/T_\nu$ and neglect terms involving $m/T_k\ll1$
\begin{align}\label{rho}
\rho^{EV}&\simeq \frac{g_\nu T_\nu^4}{2\pi^2}\int_0^\infty\frac{\left(m_\nu^2/T_\nu^2+u^2\right)^{1/2}u^2}{\Upsilon^{-1}\exp(u)+ 1}du,\\
P^{EV}&\simeq \frac{g_\nu T_\nu^4}{6\pi^2}\int_0^\infty\frac{\left(m_\nu^2/T_\nu^2+u^2\right)^{-1/2}u^4}{\Upsilon^{-1}\exp(u)+ 1}du,
\label{Pev}
\end{align}
where the upper index `EV' reminds us that we have used the Einstein-Vlasov free streaming solution for the neutrino distribution.  For $m\ll T_\nu$ the massless equation of state $\rho^{EV}=3P^{EV}$ holds, but when $T_\nu$ is on the order of the mass, the mass term becomes important and modifies the equation of state.  The lack of a mass term in the exponential gives this a distinctly different behavior from the equilibrium Fermi-Dirac distribution.

To illustrate the effect of fugacity and neutrino mass on the equation of state, we examine the energy density and pressure of the neutrino distribution.  We separate off the zero mass,  $\Upsilon_\nu=1$ contributions from a single neutrino flavor with standard reheating by defining
\begin{equation}\label{rho0}
\rho_0=\frac{7\pi^2 }{120} \left[\left(\frac{4}{11}\right)^{1/3}T_\gamma\right]^4,\hspace{2mm} P_0=\rho_0/3.
\end{equation}

We note that $\rho^{EV}/\rho_0$ and $P^{EV}/P_0$ are functions of $(m/T_\gamma)^2$, and $N_\nu$, where the latter dependence is obtained by inverting both $\Upsilon(T_\gamma/T_\nu)$ from \req{Upsilon_ratio} and  $N_\nu(\Upsilon)$ from \req{N_nu_approx} in order to obtain $T_\gamma/T_\nu(\Upsilon)$ and  $\Upsilon(N_\nu)$.  In practice, these inversions are best done numerically. Thus the quantities of interest are  $\rho^{EV}/\rho_0$ from  \req{rho} and \req{rho0} and the corresponding expressions for the pressure, both as functions of $\delta N_\nu=N_\nu-3$ and $\beta=m_\nu/T_\gamma$.  Again, we emphasize that here $N_\nu$ refers to the value of the effective number of neutrinos that was established at neutrino freeze-out when neutrinos were still effectively massless.  At the temperatures we are now considering, \req{N_nu_c} no longer applies.

The functional dependence of the energy density and pressure that we find is best characterized by a simple polynomial representation that arises from   a least squares fit:
\begin{align}\label{rho_P_fits}
&\rho^{EV}/\rho_0= N_\nu+0.1016\sum_i\beta_i^2+0.0015\delta N_\nu\sum_i\beta_i^2\notag\\
&-0.0001\delta N_\nu^2\sum_i\beta_i^2-0.0022\sum_i\beta_i^4,\\
&P^{EV}/P_0= N_\nu-0.0616\sum_i\beta_i^2-0.0049\delta N_\nu\sum_i\beta_i^2\notag\\
&+0.0005\delta N_\nu^2\sum_i\beta_i^2+0.0022\sum_i\beta_i^4.\label{tau_Ups}
\end{align}
For a single massive neutrino, these fits  are valid to within $2\%$ and $5\%$ relative error respectively in the region $3\leq N_\nu \leq 5$, $0\leq m_\nu/T_{\gamma}\leq 4$. Note that the upper limit is sufficient to cover the era of electron-ion recombination at $T_\gamma=O(0.3)$ eV and neutrino masses less than the upper bound $\sum m_i<0.23$ eV reported in \cite{Planck}.

 For small mass to temperature ratios, the fit depends  only on the sum of neutrino masses squared, $\sum_i\beta_i^2\equiv (\sum_im_i^2)/T_\gamma^2$. However, the fourth order term is not negligible over the chosen fitting region.  Removing that term results in a maximum relative error between the fit and the exact result of greater than $25\%$. This indicates that one maybe able to use a fit to cosmic data to constrain the hierarchy structure of the neutrino mass spectrum using the fitted values of $\sum_im_i^2$ and $\sum_im_i^4$.

The formulas in \req{rho_P_fits}, or the more precise quantities \req{rho} and \req{Upsilon_ratio} that we fit in order to obtain them, should be used when exploring the combined effects of $N_\nu\neq 3$ and neutrino mass on cosmological observables as they properly capture the interplay between neutrino mass and the shape of the neutrino distribution in terms of physical observables. In particular, they can be used to extract fits to $\sum_i m_i^2$ and $\sum_i m_i^4$ while separating off the effects of the confounding variable $N_\nu$.

\begin{figure}[h]
\centerline{\includegraphics[height=6.2cm]{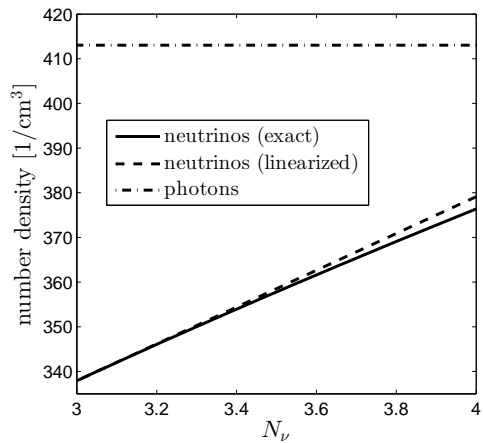}}
\caption{Total neutrino and  anti-neutrino number density in the Universe `today' as a function of $N_\nu$, both exact (solid) and linearized (dashed). The photon density (dot-dashed) is also shown as a reference.} \label{fig:neutrino_density}
 \end{figure}

Similarly we find the net number of neutrinos after freeze-out as function of the effective number of neutrinos $N_\nu$.  Using \req{num_density}, \req{Upsilon_ratio}, and \req{N_nu_approx} to Taylor expand the neutrino number density as a function of $\delta N_\nu$ we find
\begin{equation}\label{n_density_lin}
n_{\rm today}=(0.1993+0.02429\delta N_\nu)T_\gamma^3.
\end{equation}
In the  present day Universe, at $T_\gamma=T_{CMB}=0.2349$ meV, we show both the exact and linearized results in figure \ref{fig:neutrino_density}.  We note that at $\delta N_\nu=0$ we find $112.6$ cm${}^{-3}$ per flavor, in close agreement with \cite{Wong}.
%%%%%%%%%%%%%%%%%%%%%%%%%%%%%%%%%%%%%%%
%%%%%%%%%%%%%%%%%%%%%%%%%%%%%%%%%%%%%%%

%%%%%%%%%%%%%%%%%%%%%%%%%%%%%%%%%%%%%%%%%%%%%%%%%%%%%%%%%%
\section{Discussion}\label{discuss}
The cosmic neutrino momentum distribution is a critical input into our understanding of the spectrum of CMB fluctuations and arguably its  understanding is a prerequisite for the consideration of cosmic neutrino detection opportunities.  Motivated by hints of a tension between the Planck results and the standard theory of neutrino freeze-out expressed by a noticeably non-integer value of the  effective number of neutrinos $N_\nu$, we have undertaken a model independent study of the effect of a delayed kinetic freeze-out on $N_\nu$, and more generally on the form of the neutrino momentum distribution.  The search for reaction mechanisms that can produce a reduction in the neutrino freeze-out temperature  $T_k$ from a value near $T_k/m_e\simeq 1.1$  to a value perhaps as small as   $T_k/m_e\simeq 0.35$, see figure \ref{fig:N_eff}, is a topic for future investigation.

Possible participation of neutrinos in  $e^\pm$ annihilation reheating and hence a ratio of photon to neutrino temperature that is closer to one, and thus $N_\nu>3$  is  a well known fact~\cite{Mangano:2005cc}.  However,  less  appreciated is the impact of neutrino reheating on neutrino fugacity $\Upsilon_\nu$, the factor describing the   the neutrino distribution compared to chemical equilibrium  $\Upsilon_\nu=1$.  

We have shown how  an increase in $N_\nu$ is naturally interpreted to be due to a delayed neutrino kinetic freeze-out temperature $T_k$. We derived an approximate power law relation, \req{Upsilon_ratio}, between the fugacity factor and the photon to neutrino temperature ratio that arises from a delayed freeze-out.  We found a fugacity $\Upsilon_\nu$ less than unity, and thus an underpopulation of phase space compared to chemical equilibrium. 

After freeze-out, neutrinos freely stream through the expanding universe. The non-thermal modification from the fugacity factor is frozen into the shape of the distribution. We derived how this modified neutrino distribution evolves as the universe expands and explored how the energy density and pressure are modified, including in this study the interplay of fugacity and neutrino mass.  We note that these effects impact the cosmological study of the question of neutrino mass hierarchy: the effects we present in \req{rho_P_fits} and \req{tau_Ups} produce a functional dependence on both $\sum_im^2_i$ and $\sum_im^4_i$.

The fits \req{rho_P_fits} and \req{tau_Ups} of the exact, but intractable \req{rho} and \req{Pev} show how the energy density and pressure are self consistently modified when both neutrino mass and a non-integer $N_\nu$  are present.  The latter dependence is expressed by using  $T_\gamma/T_\nu(N_\nu)$ and $\Upsilon_\nu(N_\nu)$ in \req{rho} and \req{Pev} as discussed in section \ref{nu_mass_section}. This polynomial presentation of the cosmic neutrino energy density and pressure resolves an old problem of cosmic neutrino physics  by allowing  a physically consistent treatment of the combined effects of neutrino mass and $\delta N_\nu$ as long as the magnitude of $\delta N_\nu$ follows from  neutrino freeze-out dynamics.

%%%%%%%%%%%%%%%%%%%%%%%%%%%%%%%%%
\subsection*{Acknowledgments}
This work has been supported by a grant from the U.S. Department of Energy, DE-FG02-04ER41318
and by the Department of Defense (DoD) through the National Defense Science \& Engineering Graduate Fellowship (NDSEG) Program.

%%%%%%%%%%%%%%%%%%%%%%%%%%%%%%%%%%%%%%%%%%%%%%%%%%%%%%%%%%

%\bibliographystyle{unsrt}
%\bibliography{refs} 

\end{document}